\documentclass[11pt]{article}
\usepackage{graphicx}
\usepackage{amsmath}
\usepackage{amssymb}
\usepackage{caption2}
\begin{document}
\bibliographystyle{prsty}
\begin{center}
{\large {\bf \sc{  Analysis of  the $Y(4274)$ with  QCD sum rules }}} \\[2mm]
Zhi-Gang Wang \footnote{E-mail,wangzgyiti@yahoo.com.cn.  }     \\
 Department of Physics, North China Electric Power University,
Baoding 071003, P. R. China
\end{center}

\begin{abstract}
In this article, we assume that there exists a pseudoscalar
$ {\bar D}_sD_{s0}(2317)$  molecular state  and study its mass with
the molecule-type  interpolating  current in details using the QCD
sum rules. The numerical result disfavors identifying the
charmonium-like state $Y(4274)$ as the ${\bar D}_sD_{s0}(2317)$
molecule.
\end{abstract}



\section{Introduction}

In 2009, the CDF collaboration   observed   a narrow structure (the
$Y(4140)$) near the $J/\psi\phi$ threshold with a statistical
significance greater than  $3.8\,\sigma$  in exclusive $B^+\to
J/\psi\phi K^+$ decays produced in $\bar{p} p $ collisions
\cite{CDF0903}. The measured  mass and width are
$(4143.0\pm2.9\pm1.2)\,\rm{ MeV}$ and $(11.7^{+8.3}_{-5.0}\pm3.7)\,
\rm{MeV}$, respectively. There have been several identifications for
the $Y(4140)$, such as the molecular state \cite{Y4140-molecule},
the tetraquark state \cite{Y4140-tetraquark,Drenska0902}, the hybrid
state \cite{Y4140-hybrid,Y4140-Wang}, the re-scattering effect \cite{Y4140-cc},
etc. The Belle collaboration measured the process $\gamma \gamma \to
\phi J/\psi$ for the $\phi J/\psi$ invariant mass distributions
between the threshold and $5\,\rm{GeV}$, and observed no signal for
the structure $Y(4140)\to \phi J/\psi$, however, they observed a
narrow peak (the $X(4350)$) of $8.8^{+4.2}_{-3.2}$ events with an
significance of $3.2\,\sigma$ \cite{Belle4350}. The measured mass
and width  are $(4350.6^{+4.6}_{-5.1}\pm 0.7)\,\rm{MeV}$ and
$(13.3^{+17.9}_{-9.1}\pm 4.1)\,\rm{MeV}$,  respectively
\cite{Belle4350}.  Recently, the CDF collaboration confirmed the
$Y(4140)$ in the $B^\pm\rightarrow J/\psi\,\phi K^\pm$ decays  with
a  statistical significance greater  than $5\,\sigma$, the
Breit-Wigner mass and width are $\left(4143.4^{+2.9}_{-3.0} \pm0.6
\right)\, \rm{MeV}$ and
$\left(15.3^{+10.4}_{-6.1}\pm2.5\right)\,\rm{MeV}$, respectively
\cite{CDF1101}, which are consistent with the values from the
earlier CDF analysis \cite{CDF0903}. Furthermore, the CDF
collaboration  observed an evidence for a second structure (the
$Y(4274)$) with approximate significance of $3.1\,\sigma$. The
measured mass and width
 are $\left(4274.4^{+8.4}_{-6.7}\pm1.9\right)\,\rm{MeV}$ and
$\left(32.3^{+21.9}_{-15.3}\pm7.6\right)\,\rm{MeV}$, respectively
\cite{CDF1101}.

In Ref.\cite{LiuLZ1011}, Liu, Luo and Zhu  identify the
charmonium-like state $Y(4274)$ as the $S$-wave
$D_s\bar{D}_{s0}(2317)+h.c.$ molecule with the spin-parity
$J^P=0^-$, and make prediction for the mass of  the $S$-wave
$D\bar{D}_{0}(2400)+h.c.$ molecule as its cousin. In
Ref.\cite{HeLiu1102}, He and Liu take the $Y(4274)$ as the
$D_s\bar{D}_{s0}(2317)$ molecular state and study the line shapes of
the open-charm radiative decays and pionic decays. In
Ref.\cite{Finazzo1102}, Finazzo, Liu and Nielsen  study the
 $Y(4274)$ as the $D_s\bar{D}_{s0}(2317)+h.c.$
molecular state with $J^{PC}=0^{-+}$ using the QCD sum rules, and
obtain the mass $M_{D_s\bar{D}_{s0}}=(4.78\pm 0.54)\,\rm{GeV}$.

The mass is a fundamental parameter in describing a hadron, in order
to identify  the $Y(4274)$ as the $D_s\bar{D}_{s0}(2317)$ (or
$\bar{D}_sD_{s0}(2317)$) molecular state, we must prove that its
mass lies in the region $(4.2-4.3)\, \rm{GeV}$. The normal threshold
 of the scattering state $\bar{D}_sD_{s0}(2317)$  is about
$4287\,\rm{MeV}$ \cite{PDG}, which is slightly larger than the mass
of the $Y(4274)$. In this article, we assume that there exists a
pseudoscalar ${\bar D}_sD_{s0}(2317)$ molecular state in the
$J/\psi\phi$ invariant mass distribution indeed, and study its mass
using the QCD sum rules \cite{SVZ79,Reinders85}\footnote{In
preparing the article, Ref.\cite{Finazzo1102} appears. We take into
account the vacuum condensates adding up to dimension-10
consistently, while in the second version of Ref.\cite{Finazzo1102},
some vacuum condensates such as $\langle
\bar{s}s\rangle\langle\frac{\alpha_sGG}{\pi}\rangle$, $\langle
\bar{s}g_s\sigma Gs\rangle\langle\frac{\alpha_sGG}{\pi}\rangle$ are
still  neglected.}.

The article is arranged as follows:  we derive the QCD sum rules for
the mass of  the $Y(4274)$  in section 2; in section 3, numerical
results and discussions; section 4 is reserved for conclusion.

\section{QCD sum rules for  the molecular state $Y(4274)$ }
In the following, we write down  the two-point correlation function
$\Pi(p)$  in the QCD sum rules,
\begin{eqnarray}
\Pi(p)&=&i\int d^4x e^{ip \cdot x} \langle
0|T\left\{J(x)J^{\dagger}(0)\right\}|0\rangle \, , \\
J(x)&=&\bar{c}(x)i\gamma_5 s(x) \bar{s}(x) c(x) \, ,
\end{eqnarray}
we choose  the pseudoscalar current $J(x)$ to interpolate the
molecular state $Y(4274)$. There are two additional pseudoscalar currents  $\eta_5(x)$ and $J_5(x)$,
\begin{eqnarray}
\eta_5(x)&=&\bar{c}(x)\gamma_\mu \gamma_5s(x) \bar{s}(x)\gamma^\mu c(x) \, , \nonumber\\
J_5(x)&=&\bar{c}(x)\sigma_{\mu\nu}\gamma_5 s(x) \bar{s}(x)\sigma^{\mu\nu} c(x) \, ,
\end{eqnarray}
which interpolate the pseudoscalar
   $D^*_s(2112) \bar{D}^*_{s1}(2460)$ and $D_{s1}(2536)\bar{D}_{s1}(2536)$ molecular states, respectively,
here we have neglected the
 mixing between the   $D^*_{s1}(2460)$ and $D_{s1}(2536)$ \cite{Mixing}, and take them as the ${}^3P_1$ and ${}^1P_1$ states respectively in
 the non-relativistic quark model considering the masses $M_{D^*_{s1}}\approx M_{D_{s1}}$. The normal thresholds
 of the scattering states $\bar{D}_sD_{s0}(2317)$, $D^*_s(2112)\bar{D}^*_{s1}(2460)$ and $D_{s1}(2536)\bar{D}_{s1}(2536)$ are
$4287\,\rm{MeV}$, $4572\,\rm{MeV}$ and $5070\,\rm{MeV}$, respectively \cite{PDG}. It is impossible  for the
$D_{s1}(2536)\bar{D}_{s1}(2536)$ molecular state
has the binding energy $-796\,\rm{MeV}$ so as to reproduce the mass of the $Y(4274)$. In Ref.\cite{Y4140-Wang},
we perform detailed studies of the
$D_s^\ast(2112) {\bar D}_s^\ast(2112)$ molecular state using the interpolating  current,
\begin{eqnarray}
\eta(x)&=&\bar{c}(x)\gamma_\mu s(x) \bar{s}(x)\gamma^\mu c(x) \, ,
\end{eqnarray}
and obtain the  value $M_{D_s^\ast {\bar
D}_s^\ast}=(4.48\pm0.17)\,\rm{GeV}$ or $(4.43\pm0.16)\,
\rm{GeV}$, which is larger than the normal threshold
 of the scattering state $D_s^*(2112)\bar{D}_{s}^*(2112)$, and draw the conclusion tentatively that the $D_s^*(2112)\bar{D}_{s}^*(2112)$ is probably virtual
state not related with the $Y(4140)$. Compared with the scalar current $\eta(x)$, the pseudoscalar current $\eta_5(x)$ has an additional Dirac matrix
$\gamma_5$, which leads  to
the terms $m_c^2\langle\bar{s}s\rangle^2$, $m_c^2\langle\bar{s}g_s \sigma Gs\rangle^2$, and some terms $m_c\langle\bar{s}s\rangle$, $m_c\langle\bar{s}g_s \sigma Gs\rangle$ change their sign, and the convergent behavior in the operator product expansion
  becomes  worse, we have to choose much larger Borel parameter and postpone the threshold parameter
to much larger value, and obtain the value $M_{D_s^\ast {\bar
D}_{s1}^\ast}$, which is much larger than the $M_{D_s^\ast {\bar
D}_s^\ast}$. It is difficult to reproduce the mass of the $Y(4274)$ as the  pseudoscalar
   $D^*_s(2112)\bar{D}^*_{s1}(2460)$   molecular state using the QCD sum rules.

We can insert  a complete set of intermediate hadronic states with
the same quantum numbers as the current operator $J(x)$ into the
correlation function $\Pi(p)$  to obtain the hadronic representation
\cite{SVZ79,Reinders85}. After isolating the ground state
contribution from the pole term, which is supposed to be the
$Y(4274)$, we get the following result,
\begin{eqnarray}
\Pi(p)&=&\frac{\lambda_{Y}^2}{M_{Y}^2-p^2} +\cdots \, \, ,
\end{eqnarray}
where the pole residue (or coupling) $\lambda_Y$ is defined by
\begin{eqnarray}
\lambda_{Y} &=& \langle 0|J(0)|Y(p)\rangle \, .
\end{eqnarray}

The  contributions from the two-particle and many-particle reducible
states are  small enough to be neglected \cite{Lee2005}.  The
scattering state $\bar{D}_s D_{s0}(2317)$ has the same quantum
numbers as the current operator $J(x)$, the corresponding
 two-particle reducible contribution can be written as
\begin{eqnarray}
\Pi_{2}(p)&=&i\lambda_{\bar{D}_s D_{s0}}^2\int~{d^4q\over(2\pi)^4}
\frac{1}{\left[ q^2-m_{D_s}^2\right]\left[
(p-q)^2-m_{D_{s0}}^2\right]}+\cdots \, ,
\end{eqnarray}
where the pole residue (or coupling) $\lambda_{\bar{D}_s D_{s0}}$
is defined by
$\langle0|J(0)|\bar{D}_s D_{s0}(p)\rangle=\lambda_{\bar{D}_s D_{s0}}
$. In the soft $D_s$ meson limit, the pole residue can be estimated
as
\begin{eqnarray}
\langle0|J(0)|\bar{D}_s D_{s0}(p)\rangle&=&-\frac{i}{f_{D_s}}\langle0|\left[Q_5,J(0)\right]|D_{s0}(p)\rangle\,,\nonumber\\
Q_5&=&\int d^3x\bar{s}(x)\gamma_0\gamma_5c(x)\,,
\end{eqnarray}
 the $f_{D_s}$ is the decay constant of the
 $D_s$ meson. We can carry out the commutator in Eq.(8) and obtain a current
consists of four quarks. The coupling  of the scalar meson  $D_{s0}(2370)$  with a four-quark current
should be very small \cite{Lee2005,Nielsen0907}. In this article, we
take the $D_{s0}(2370)$ as the conventional two-quark meson.

The $J(x)$ maybe also have nonvanishing couplings with the following
two-particle scattering states, such as the $\chi_{c0}\eta(548)$,
$\chi_{c0}\eta^{\prime}(958)$, $J/\psi\eta(548)$,
$J/\psi\eta^{\prime}(958)$, ${J/\psi}f_1^{\prime}(1510)$,
 $h_ch_1(1380)$,  $\chi_{c1}\phi(1020)$,
$\eta_{c}\phi(1020)$,  $\eta_cf_0^{\prime}(1500)$. The corresponding
normal thresholds are $M_{\chi_{c0}\eta}=3962.6\,\rm{MeV}$,
$M_{\chi_{c0}\eta^{\prime}}=4372.53\,\rm{MeV}$,
$M_{J/\psi\eta}=3644.77\,\rm{MeV}$,
$M_{J/\psi\eta^{\prime}}=4054.7\,\rm{MeV}$,
$M_{{J/\psi}f_1^{\prime}}=4614.92\,\rm{MeV}$,
$M_{h_ch_1}=4911.42\,\rm{MeV}$,
$M_{\chi_{c1}\phi}=4530.12\,\rm{MeV}$,
$M_{\eta_{c}\phi}=3999.76\,\rm{MeV}$ and
$M_{\eta_cf_0^{\prime}}=4485.3\,\rm{MeV}$, respectively \cite{PDG}.
We can write the interpolating  current $J(x)$ in the following form
with the Fierz re-ordering,
\begin{eqnarray}
J(x)&=&-\frac{1}{12}\bar{c}(x)c(x)\bar{s}(x)i\gamma_{5}s(x)
-\frac{i}{12}\bar{c}(x)\gamma_{\alpha}c(x)\bar{s}(x)\gamma^{\alpha}\gamma_{5}s(x)
\nonumber\\
&&-\frac{i}{24}\bar{c}(x)\sigma_{\alpha\beta}c(x)\bar{s}(x)\sigma^{\alpha\beta}\gamma_{5}s(x)
+\frac{i}{12}\bar{c}(x)\gamma_{\alpha}\gamma_5c(x)\bar{s}(x)\gamma^{\alpha}s(x)\nonumber\\
&&-\frac{1}{12}\bar{c}(x)i\gamma_{5}c(x)\bar{s}(x)s(x)-\frac{1}{2}\bar{c}(x)\frac{\lambda^a}{2}c(x)\bar{s}(x)\frac{\lambda^a}{2}i\gamma_{5}s(x)
\nonumber\\
&&-\frac{i}{2}\bar{c}(x)\frac{\lambda^a}{2}\gamma_{\alpha}c(x)\bar{s}(x)\frac{\lambda^a}{2}\gamma^{\alpha}\gamma_{5}s(x)
-\frac{i}{4}\bar{c}(x)\frac{\lambda^a}{2}\sigma_{\alpha\beta}c(x)\bar{s}(x)\frac{\lambda^a}{2}\sigma^{\alpha\beta}\gamma_{5}s(x)\nonumber\\
&&+\frac{i}{2}\bar{c}(x)\frac{\lambda^a}{2}\gamma_{\alpha}\gamma_5c(x)\bar{s}(x)\frac{\lambda^a}{2}\gamma^{\alpha}s(x)
-\frac{1}{2}\bar{c}(x)\frac{\lambda^a}{2}i\gamma_{5}c(x)\bar{s}(x)\frac{\lambda^a}{2}s(x)\,,\nonumber\\
&=&-\frac{1}{12}J_{SP}^0(x) -\frac{i}{12}J_{VA}^0(x)
-\frac{i}{24}J_{TT}^0(x)
+\frac{i}{12}J_{AV}^0(x)-\frac{1}{12}J_{PS}^0(x)\nonumber\\
&&-\frac{1}{2}J^8_{SP}(x)-\frac{i}{2}J^8_{VA}(x)
-\frac{i}{4}J^8_{TT}(x)+\frac{i}{2}J^{8}_{AV}(x)
-\frac{1}{2}J^{8}_{PS}(x)\,,
\end{eqnarray}
where the $J^0_{k}$ and $J^8_{k}$ are the color-singlet and
color-octet currents, respectively. Here we have used the following identity,
\begin{eqnarray}
\delta_{ij}\delta_{mn}&=&\frac{1}{3}\delta_{in}\delta_{mj}+2\left(\frac{\lambda^a}{2}\right)_{in}\left(\frac{\lambda^a}{2}\right)_{mj}\,,
\end{eqnarray}
to perform the re-arrangement in the color space. There is another identity,
\begin{eqnarray}
\delta_{ij}\delta_{mn}&=&\frac{9}{4}\left(\frac{\lambda^a}{2}\right)_{in}\left(\frac{\lambda^a}{2}\right)_{mj}+\frac{3}{4}\left(\frac{\lambda^a}{2}\right)_{ij}\left(\frac{\lambda^a}{2}\right)_{mn}\,,
\end{eqnarray}
to express the current $J(x)$ into  a series of  color-octet currents, the identities in Eqs.(10-11) can be changed into
  each other.
 The couplings to the
lowest scattering states below the normal $\bar{D}_s D_{s0}(2317)$
threshold can be estimated as
\begin{eqnarray}
\langle0|J^0_{SP}(0)|\chi_{c0}\eta(p)\rangle&=&-\frac{i}{f_{\eta}}\langle0|\left[Q_5,J^0_{SP}(0)\right]|\chi_{c0}(p)\rangle\, ,\nonumber\\
\langle0|J^0_{VA}(0)|J/\psi\eta/\eta^{\prime}(p)\rangle&=&-\frac{i}{f_{\eta/\eta^{\prime}}}\langle0|\left[Q_5,J^0_{VA}(0)\right]|J/\psi(p)\rangle\, ,\nonumber\\
\langle0|J^0_{AV}(0)|\eta_{c}\phi(p)\rangle&=&-\frac{i}{f_{\eta_{c}}}\langle0|\left[Q_5,J^0_{AV}(0)\right]|\phi(p)\rangle\,,
\end{eqnarray}
in the soft  pseudoscalar mesons limit, where the $Q_5$ are the
axial-charges  in the due channels, the $f_{\eta}$,
$f_{\eta^{\prime}}$ and $f_{\eta_c}$ are the decay constants of the
pseudoscalar mesons. If we carry out the calculations for the commutators in Eq.(12), we obtain the currents
consist of four quarks.  The couplings of the four-quark currents to the
two-quark mesons are supposed to be small, the contaminations from
the two-particle reducible scattering states can be neglected.

On the other hand, the pseudoscalar charmonia $\eta_c$, $\eta_c'$,
$\eta_c''$, $\cdots$  have Fock states with additional $q\bar{q}$
components besides
 the $c\bar{c}$ components. The current $J(x)$  maybe have nonvanishing couplings
 with the pseudoscalar charmonia, those couplings are supposed to be small, as the
 dominating  Fock states of the pseudoscalar charmonia
  are the $c\bar{c}$ components. The contaminations from the
  conventional charmonia can also be  neglected.

The molecular current $J(x)$ can also be expressed in terms of  a series of diquark-antidiquark type currents with the
rearrangement  in the Dirac spinor space and
color space, in other words, the $\bar{c}(x)\Gamma c(x)\bar{s}(x) \Gamma' s(x)$ type currents in the $J(x)$ can be recasted
   into
 a special superposition of the  diquark-antidiquark type currents themselves.
 The molecular current $\bar{c}(x)\Gamma c(x)\bar{s}(x) \Gamma' s(x)$
 maybe have non-vanishing couplings  with  a series of
  tetraquark states, we cannot distinguish those
contributions  to study them exclusively, and   assume that the  current $\bar{c}(x)\Gamma c(x)\bar{s}(x) \Gamma' s(x)$
 couples to a particular resonance, the molecular state (for example, the $\chi_{c0}\eta$, $J/\psi\eta$, $\eta_{c}\phi$, etc),
  which is a special  superposition of the
tetraquark states (or has the tetraquark states as its Fock components), and embodies   the  net
effects. On the other hand, one can resort to the tetraquark scenario instead of the molecule scenario to study the possible contaminations.
However, it is a very hard work, as the predicted masses of the tetraquark states from different theoretical approaches differ from each other greatly, and
the tetraquark states have not been confirmed experimentally.

 In the following, we briefly outline  the operator product
expansion for the correlation function $\Pi(p)$  in perturbative
QCD.  We contract the quark fields in the correlation function
$\Pi(p)$ with Wick theorem, obtain the result:
\begin{eqnarray}
\Pi(p)&=&i\int d^4x e^{ip \cdot x}   Tr\left[i\gamma_5
S_{ab}(x)i\gamma_5 C_{ba}(-x) \right] Tr\left[ C_{mn}(x) S_{nm}(-x)
\right] \, ,
\end{eqnarray}
where the $a$, $b$, $m$ and $n$ are color indexes, and  substitute
the full  $s$ and $c$ quark propagators $S_{ab}(x)$ and $C_{ab}(x)$
into the correlation function $\Pi(p)$ and complete  the integral in
the coordinate space, then integrate over the variables in the
momentum space,  and obtain the correlation function $\Pi(p)$ at the
level of the quark-gluon degrees  of freedom.

 Once analytical results are obtained,   then we can take the
quark-hadron duality and perform Borel transform  with respect to
the variable $P^2=-p^2$, finally we obtain  the following sum rule:
\begin{eqnarray}
\lambda_{Y}^2 e^{-\frac{M_Y^2}{M^2}}= \int_{4(m_c+m_s)^2}^{s_0} ds
\rho(s) e^{-\frac{s}{M^2}} \, ,
\end{eqnarray}
where
\begin{eqnarray}
\rho(s)&=&\rho_{0}(s)+\rho_{\langle\bar{s}s\rangle}(s)+\rho_{\langle
\bar{s}s\rangle^2}(s)+\left[\rho^A_{\langle
GG\rangle}(s)+\rho^B_{\langle GG\rangle}(s)\right]\langle
\frac{\alpha_s GG}{\pi}\rangle\nonumber\\
&&+\rho_{\langle GGG\rangle}(s) \langle g_s^3GGG\rangle\, ,
\end{eqnarray}
the lengthy  expressions of the spectral densities $\rho_0(s)$,
$\rho_{\langle \bar{s}s\rangle}(s)$, $\rho_{\langle
\bar{s}s\rangle^2}(s)$, $\rho^A_{\langle GG\rangle}(s)$,
$\rho^B_{\langle GG\rangle}(s)$  and $\rho_{\langle GGG\rangle}(s)$
are presented in the appendix. In this article, we carry out the
operator product expansion to the vacuum condensates adding up to
dimension-10 and
 take the assumption of vacuum saturation for the  high
dimension vacuum condensates.

 Differentiate   Eq.(14) with respect to  $\frac{1}{M^2}$, then eliminate the
 pole residue $\lambda_{Y}$, we can obtain a sum rule for
 the mass of the $Y(4274)$,
 \begin{eqnarray}
 M_Y^2= \frac{\int_{4(m_c+m_s)^2}^{s_0} ds
\frac{d}{d \left(-1/M^2\right)}\rho(s)e^{-\frac{s}{M^2}}
}{\int_{4(m_c+m_s)^2}^{s_0} ds \rho(s)e^{-\frac{s}{M^2}}}\, .
\end{eqnarray}

\section{Numerical results and discussions}
The input parameters are taken to be the standard values $\langle
\bar{q}q \rangle=-(0.24\pm 0.01\, \rm{GeV})^3$, $\langle \bar{s}s
\rangle=(0.8\pm 0.2)\langle \bar{q}q \rangle$, $\langle
\bar{s}g_s\sigma G s \rangle=m_0^2\langle \bar{s}s \rangle$,
$m_0^2=(0.8 \pm 0.2)\,\rm{GeV}^2$, $\langle \frac{\alpha_s
GG}{\pi}\rangle=(0.33\,\rm{GeV})^4 $, $\langle g_s^3G G G
\rangle=0.045\,\rm{GeV}^6$, $m_s=(0.14\pm0.01)\,\rm{GeV}$
 and $m_c=(1.35\pm0.10)\,\rm{GeV}$ at the energy scale  $\mu=1\, \rm{GeV}$
\cite{SVZ79,Reinders85,Ioffe2005,ColangeloReview}.

In the conventional QCD sum rules \cite{SVZ79,Reinders85}, there are
two criteria (pole dominance and convergence of the operator product
expansion) for choosing  the Borel parameter $M^2$ and threshold
parameter $s_0$. We impose the two criteria on the
$\bar{D}_sD_{s0}(2317)$  molecular state to choose the Borel
parameter $M^2$ and threshold parameter $s_0$.

\begin{figure}
 \centering
 \includegraphics[totalheight=5cm,width=6cm]{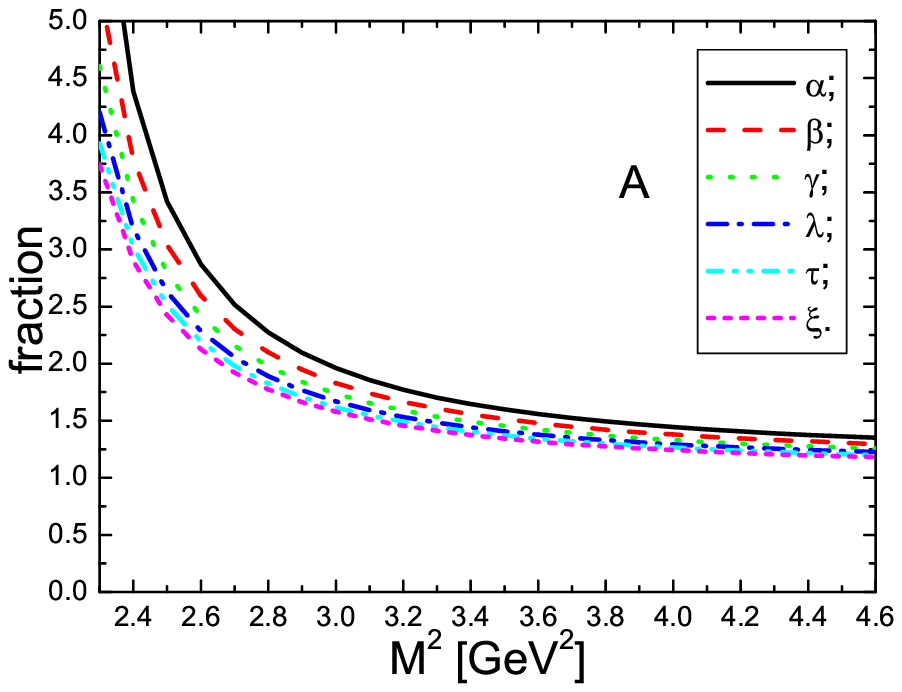}
 \includegraphics[totalheight=5cm,width=6cm]{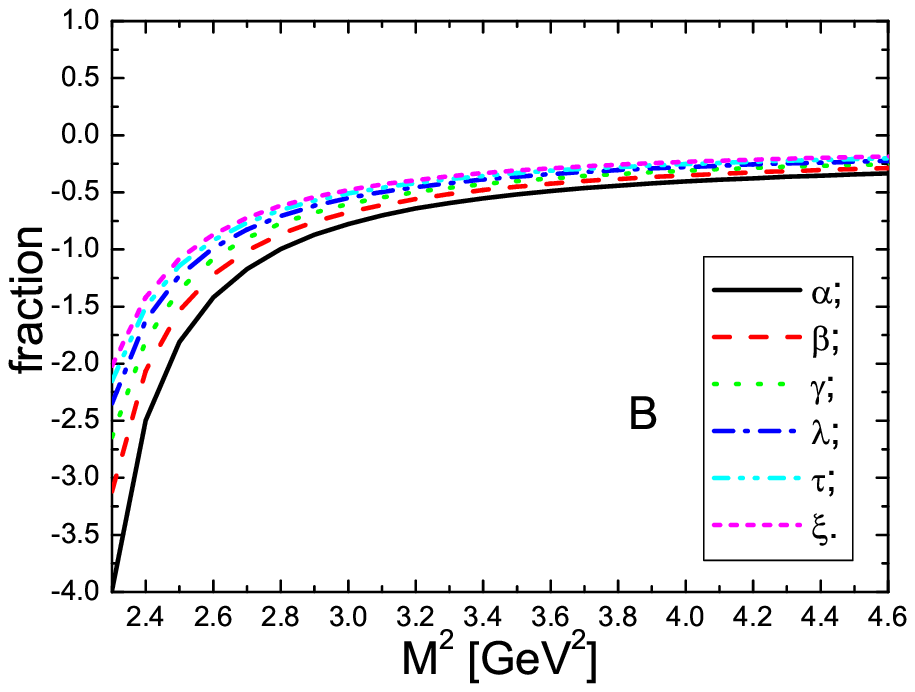}
  \includegraphics[totalheight=5cm,width=6cm]{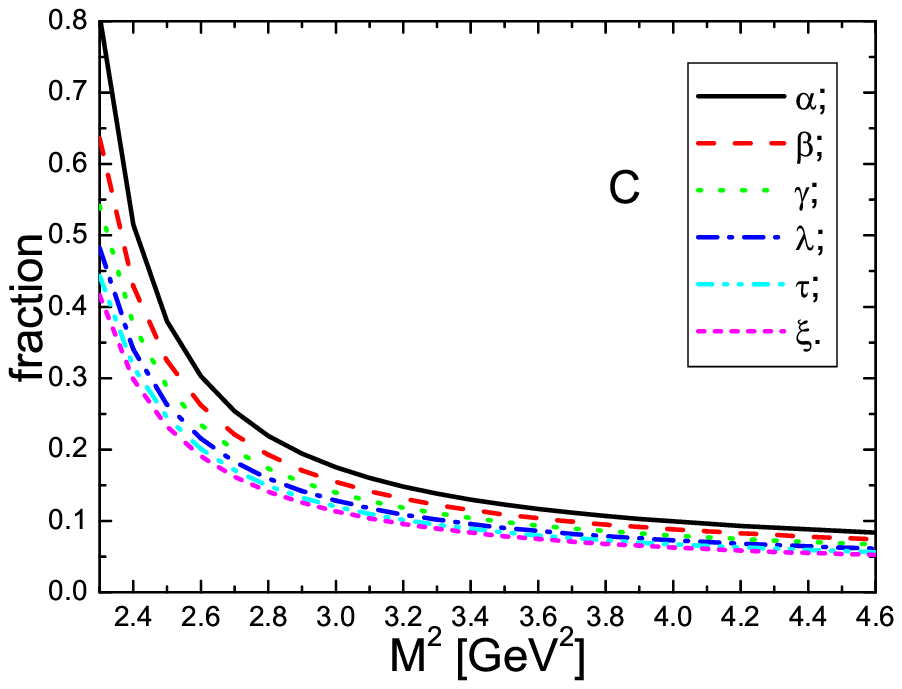}
  \includegraphics[totalheight=5cm,width=6cm]{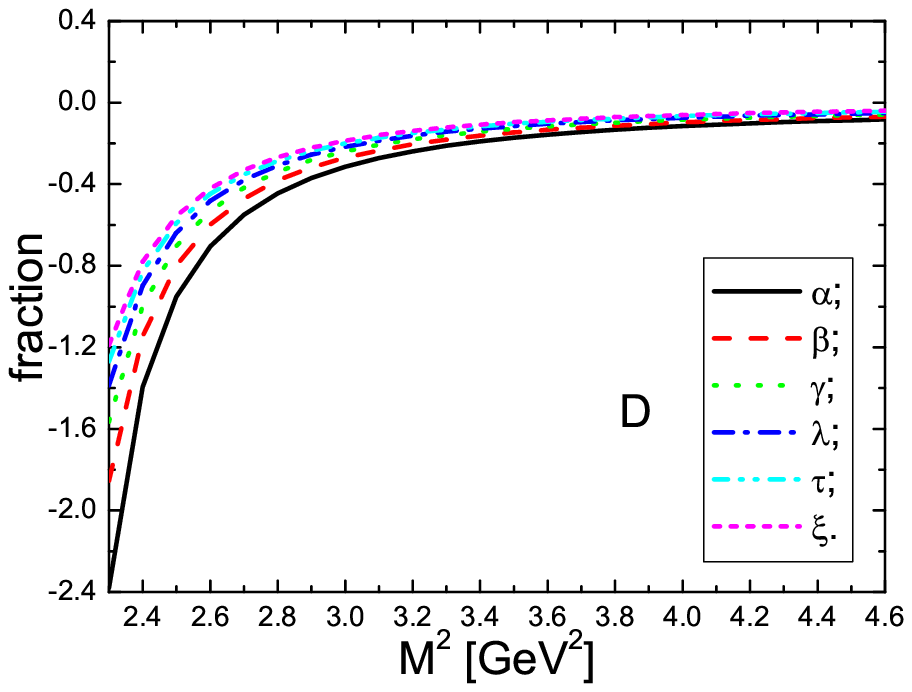}
  \includegraphics[totalheight=5cm,width=6cm]{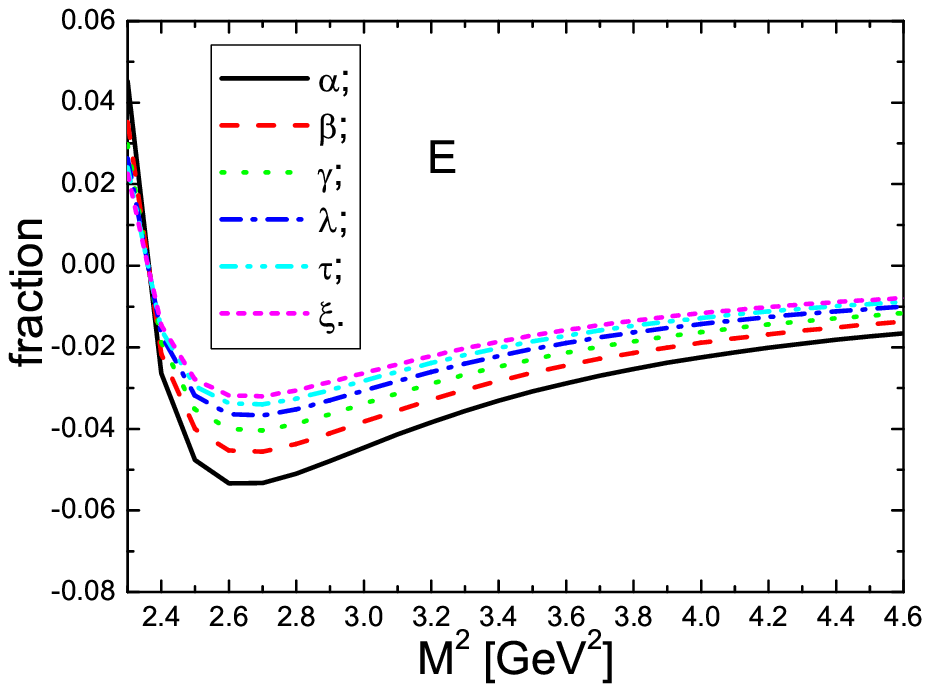}
   \includegraphics[totalheight=5cm,width=6cm]{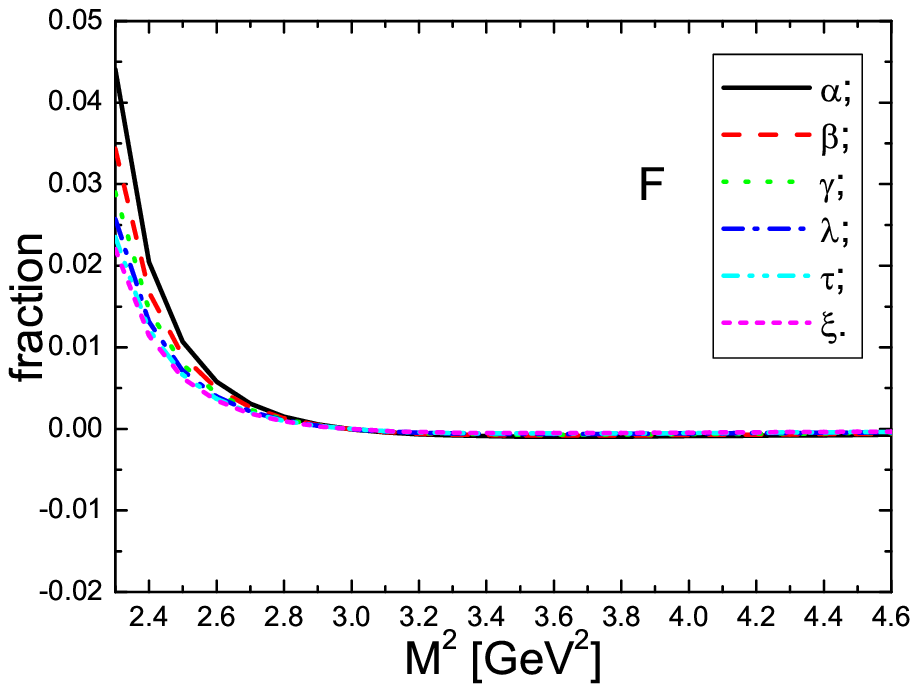}
     \caption{ The contributions from the different terms  with variations of the Borel
     parameter $M^2$ in the operator product expansion. The $A$,
   $B$, $C$, $D$, $E$ and $F$ correspond to the contributions from
   the perturbative term,
$\langle \bar{s} s \rangle+\langle \bar{s}g_s\sigma G s \rangle$
term,  $\langle \frac{\alpha_s GG}{\pi} \rangle $ + $\langle
\frac{\alpha_s GG}{\pi} \rangle \left[\langle \bar{s} s \rangle
+\langle \bar{s}g_s\sigma G s \rangle+ \langle \bar{s}s
\rangle^2\right]$ + $\langle g_s^3GGG\rangle$ term, $\langle \bar{s}
s \rangle^2$ term, $\langle \bar{s} s \rangle\langle
\bar{s}g_s\sigma G s \rangle$ term and $\langle \bar{s}g_s\sigma G s
\rangle^2$ term, respectively.     The notations
   $\alpha$, $\beta$, $\gamma$, $\lambda$, $\tau$ and $\xi$ correspond to the threshold
   parameters $s_0=26\,\rm{GeV}^2$,
   $27\,\rm{GeV}^2$, $28\,\rm{GeV}^2$, $29\,\rm{GeV}^2$, $30\,\rm{GeV}^2$ and $31\,\rm{GeV}^2$, respectively.
    Here we take the central values of the input parameters. }
\end{figure}

In Fig.1, we plot the contributions from different terms in the
operator product expansion. From the figure, we can see that the
contributions  change quickly with variations of the Borel parameter
at the region $M^2< 3\,\rm{GeV}^2$, which does  not warrant a
platform for the mass, see Fig.2. In this article, we can take the
value $M^2\geq 3\,\rm{GeV}^2$ tentatively, and the convergent
behavior in the operator product  expansion is very good.

\begin{figure}
\centering
\includegraphics[totalheight=5cm,width=6cm]{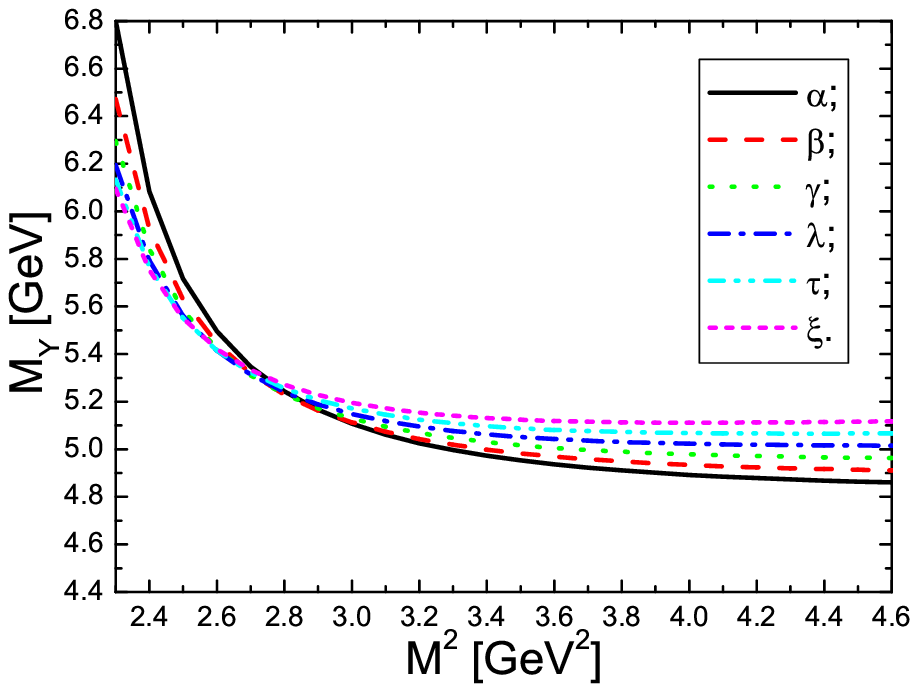}
  \caption{ The mass  with variations of the  Borel parameter $M^2$ and threshold parameter $s_0$. The notations
   $\alpha$, $\beta$, $\gamma$, $\lambda$, $\tau$ and $\xi$ correspond to the threshold
   parameters $s_0=26\,\rm{GeV}^2$,
   $27\,\rm{GeV}^2$, $28\,\rm{GeV}^2$, $29\,\rm{GeV}^2$, $30\,\rm{GeV}^2$ and $31\,\rm{GeV}^2$, respectively.
    Here we take the central values of the input parameters.   }
\end{figure}

In Fig.3, we plot the contribution from the pole term with
variations of the threshold parameter $s_0$. From the figure, we can
see that the value $s_0\leq 27 \, \rm{GeV}^2$ is too small to
satisfy the pole dominance condition. If we take the values
$s_0=(28-30)\,\rm{GeV}^2$ and $M^2=(3.0-3.6)\,\rm{GeV}^2$, the pole
contribution is about $(44-68)\%$, the pole dominance condition is
well satisfied. The Borel window changes with  variations of the
threshold parameter $s_0$, in this article, the Borel window  is
taken as $0.6\,\rm{GeV}^2$, which is small enough.    If we take
larger threshold parameter,  the Borel window is larger and the
resulting mass is larger, see Fig.2.    In this article, we intend to
obtain  the possibly  lowest mass which is supposed to be the ground
state mass  by imposing the two criteria of the QCD sum rules.
In the Borel window $M^2=(3.0-3.6)\,\rm{GeV}^2$, the
main contributions come from the perturbative term $+$ the $\langle\bar{s}s\rangle$ term $+$ the $\langle\bar{s}g_s\sigma Gs\rangle$ term, while
the dominating one is  the perturbative term,
   the convergent behavior in the operator product expansion is very good, see Fig.1.

\begin{figure}
\centering
\includegraphics[totalheight=5cm,width=6cm]{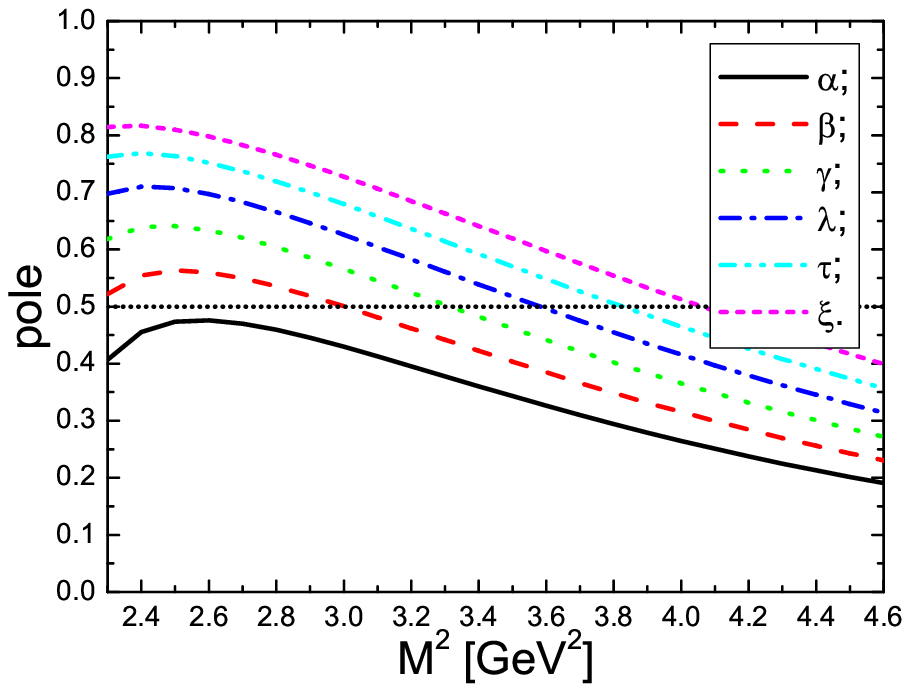}
  \caption{ The contribution from the pole term  with variations of the  Borel parameter $M^2$ and
  threshold parameter $s_0$. The notations
   $\alpha$, $\beta$, $\gamma$, $\lambda$, $\tau$ and $\xi$ correspond to the threshold
   parameters $s_0=26\,\rm{GeV}^2$,
   $27\,\rm{GeV}^2$, $28\,\rm{GeV}^2$, $29\,\rm{GeV}^2$, $30\,\rm{GeV}^2$ and $31\,\rm{GeV}^2$, respectively.
    Here we take the central values of the input parameters.   }
\end{figure}

Taking into account all uncertainties of the input parameters,
finally we obtain the values of the mass and pole residue of
 the   $Y(4274)$, which are  shown in Figs.4-5,
\begin{eqnarray}
M_{Y}&=&5.08^{+0.57}_{-0.24}\,\rm{GeV} \, ,  \nonumber\\
\lambda_{Y}&=&5.68^{+6.05}_{-1.39}\times 10^{-2}\,\rm{GeV}^5 \,   .
\end{eqnarray}
From Fig.5, we can see that at the value $M^2=3\,\rm{GeV}^2$, the
uncertainty of the pole residue is too large, the platform is not
flat enough, we can take a smaller Borel window,
$M^2=(3.2-3.6)\,\rm{GeV}^2$, then
\begin{eqnarray}
M_{Y}&=&5.06^{+0.41}_{-0.22}\,\rm{GeV} \, ,  \nonumber\\
\lambda_{Y}&=&5.56^{+3.02}_{-1.27}\times 10^{-2}\,\rm{GeV}^5 \,   .
\end{eqnarray}

In the QCD sum rules, the  high dimension vacuum condensates are
always  factorized to lower condensates with vacuum saturation,
  factorization works well in  large $N_c$ limit. In the real world,
  $N_c=3$, there are deviations from the factorable formula, we can
  introduce a factor
  $\kappa$ to parameterize the deviations,
  \begin{eqnarray}
\langle \bar{s} s \rangle^2 \, , \, \langle \bar{s} s \rangle
\langle \bar{s}g_s\sigma G s \rangle \, , \,  \langle
\bar{s}g_s\sigma G s \rangle^2 &\rightarrow& \kappa\langle \bar{s} s
\rangle^2 \, , \, \kappa\langle \bar{s} s \rangle \langle
\bar{s}g_s\sigma G s \rangle \, , \,  \kappa\langle \bar{s}g_s\sigma
G s \rangle^2 \, .
  \end{eqnarray}

In Fig.6, we plot the mass $M_Y$ with  variations of the parameter
$\kappa$ at the interval $\kappa=0-2$. From the figure, we can see
that the value of the $M_Y$ changes quickly at the region $M^2\le
3.0\,\rm{GeV}^2$, and  increases with  the $\kappa$ monotonously. At
the interval $M^2=(3.2-3.6)\,\rm{GeV}^2$, the value $\kappa=1\pm1$
leads  to an uncertainty about ${}^{+0.32}_{-0.17}\,\rm{GeV}$, which
is too small to smear the discrepancy between the present prediction
($5.06\,\rm{GeV}$) and the experimental data ($4.274\,\rm{GeV}$)
 \cite{CDF1101}.
   In the QCD sum rules for the masses of
the $\rho$ meson and
 nucleon,  $\kappa\geq 1$ \cite{Leinweber97}. If the same value holds
   for the   molecular states, the deviation from the factorable
   formula means even larger discrepancy between the present
prediction and the experimental data.

The central value $M_{Y}=5.06\,\rm{GeV}$ is about $770\,\rm{MeV}$
above the $\bar{D}_sD_{s0}(2317)$ threshold \cite{PDG}, the $\bar{D}_sD_{s0}(2317)$ is probably  a virtual state and not related to
the charmonium-like state $Y(4274)$. The present prediction is
considerably larger than the value $M_{\bar{D}_s{D}_{s0}}=(4.78\pm
0.54)\,\rm{GeV}$ \cite{Finazzo1102}.
 In Ref.\cite{Finazzo1102}, Finazzo, Liu and Nielsen take the threshold parameter
as $s_0=(M_{Y}+0.5\,\rm{GeV})^2$, and
 adjust  the threshold parameter and Borel parameter to obtain
 the Borel window and reproduce the relation
$s_0=\left(M_{Y}+(0.4\sim0.6)\,\rm{GeV}\right)^2$. While in the
present work, we search for the threshold parameter and Borel
parameter by imposing the two criteria of the QCD sum rules, and try
to obtain the ground state mass (which is not necessarily the same
as $M_Y$) in the Borel window, as the tetraquark states,
irrespective of the molecule type and the diquark-antidiquark type,
have not been firmly established yet, and we have no confidence to
take the ground state as the $Y(4274)$.

The flux-tube model \cite{FluxTube},
 the lattice QCD \cite{Latt} and the QCD string model \cite{QCD-string}
  predict that the
masses of the low lying hybrid charmonia  are about $(4.0-4.2)\,\rm{
GeV }$,  $(4.0-4.4) \, \rm{GeV}$ and $(4.2-4.5)\,\rm{GeV}$,
respectively, which are consistent with the experimental data. While
the QCD sum rules indicate that the masses of the ground-state
hybrid charmonia   with  $J^{PC}=0^{++}$, $0^{--}$ and $1^{+-}$ are
$5.4\,\rm{GeV}$, $5.8\,\rm{GeV}$ and $4.3\,\rm{GeV}$, respectively
\cite{ccG}, which disfavors identifying the $Y(4274)$ as the
$0^{++}$ hybrid charmonium because the CDF collaboration fitted the
experimental data to an $S$-wave Breit-Wigner resonance
\cite{CDF1101}.

The hybrid mesons usually decay to an $S$-wave and a $P$-wave meson
pair, and the couplings to two $S$-wave mesons are suppressed
\cite{Page1996}. If the $Y(4274)$ is a hybrid charmonium,  the decay
$Y(4274) \to J/\psi\phi$ can take place  through the final-state
re-scattering mechanism, $Y(4274)\to D_{s}\bar{D}_{s0}(2317) \to
J/\psi\phi$, and the decay  to two photons should be forbidden or
very small \cite{Page1997},
 which is in  contrary to the $X(4350)$ observed by the  Belle collaboration in
  the $\phi J/\psi$ invariant mass
distributions in the process $\gamma \gamma \to \phi J/\psi$
\cite{Belle4350}. The $X(4350)$ has been tentatively identified as
the  $2^{++}$ $cs\overline{cs}$ tetraquark state
\cite{Y4140-tetraquark,Belle4350}, the (or not the) $D_s^*\bar{D}_{s0}$
molecular state (\cite{X4350-Nielsen})
\cite{Belle4350,X4350-Zhang,X4350-Ma} , the $P$-wave charmonium
$\chi_{c2}^{\prime\prime}$ \cite{X4350-Liu}, the scalar
$cs\overline{cs}$ tetraquark state \cite{X4350-Wang,X4350-Abud}, the
scalar $\bar{c}c-{D}_s^\ast {\bar {D}}_{s}^\ast$ mixing state
\cite{X4350-Wang}, etc.

 In Ref.\cite{WangScalar},  we
study the mass spectrum of the scalar hidden charmed and hidden
bottom tetraquark states in a systematic way using the QCD sum
rules, and observe that the scalar-scalar type and
axial-vector-axial-vector type scalar $cs\bar{c}\bar{s}$ tetraquark
states have the lowest masses, about $(4.44\pm0.16)\,\rm{GeV}$,
while the pseudoscalar $cs\bar{c}\bar{s}$ tetraquark states have
much larger masses. The $Y(4274)$ may  also be an scalar
$cs\bar{c}\bar{s}$ tetraquark state, and decay $Y(4274) \to
J/\psi\phi$ takes place through fall-apart mechanism with
rearrangement in the color space.  More experimental data are still
needed to identify the new charmonium-like states.

\begin{figure}
\centering
\includegraphics[totalheight=5cm,width=6cm]{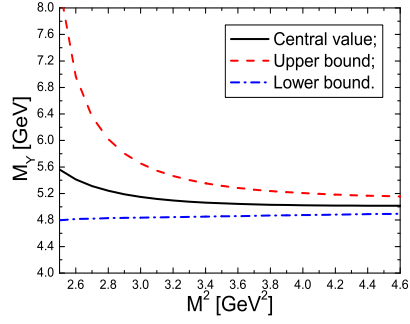}
  \caption{ The mass  with variations of the  Borel parameter $M^2$. }
\end{figure}

\begin{figure}
\centering
\includegraphics[totalheight=5cm,width=6cm]{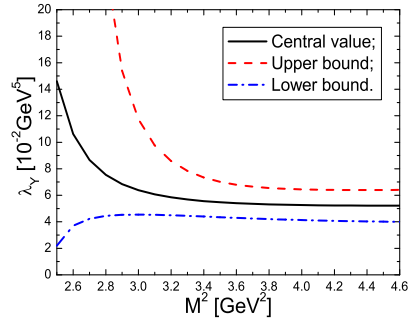}
  \caption{ The pole residue with variations of the  Borel parameter $M^2$.    }
\end{figure}

\begin{figure}
\centering
\includegraphics[totalheight=5cm,width=6cm]{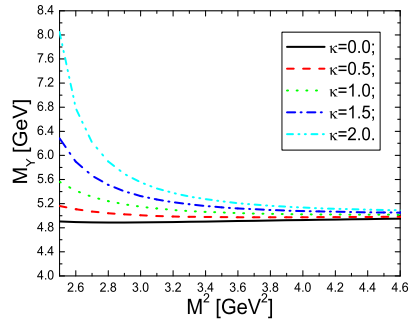}
  \caption{ The  mass with variations of the   parameters $\kappa$ and $M^2$,  other parameters are taken to be  the central values.    }
\end{figure}

\section{Conclusion}
In this article, we assume that there exists a pseudoscalar $\bar{D}_s D_{s0}(2317)$
 (or $D_s{\bar D}_{s0}(2317)$) molecular
state in the $J/\psi \phi$ invariant mass distribution indeed, and
study its mass with the  molecule-type  interpolating current in
details using the QCD sum rules.  The numerical result indicates
that the mass is about $M_{Y}=5.06^{+0.41}_{-0.22}\,\rm{GeV}$, which
 is  inconsistent with the experimental data.
    The $\bar{D}_s D_{s0}(2317)$
 (or $D_s{\bar D}_{s0}(2317)$)  is probably
    a virtual state and not related to the
charmonium-like state $Y(4274)$, and we cannot draw a definite
conclusion  with the QCD sum rules alone.

\section*{Acknowledgements}
This  work is supported by National Natural Science Foundation,
Grant Number 11075053,  and the Fundamental Research Funds for the
Central Universities.

\section*{Appendix}
The spectral densities at the level of the quark-gluon degrees of
freedom:
\begin{eqnarray}
\rho_0(s)&=&\frac{3}{2048 \pi^6}
\int_{\alpha_{i}}^{\alpha_{f}}d\alpha \int_{\beta_{i}}^{1-\alpha}
d\beta
\alpha\beta(1-\alpha-\beta)^3(s-\widetilde{m}^2_c)^2(7s^2-6s\widetilde{m}^2_c+\widetilde{m}^4_c)\,
,
\end{eqnarray}

\begin{eqnarray}
\rho_{\langle\bar{s}s\rangle}(s)&=&\frac{3m_s\langle\bar{s}s\rangle}{64
\pi^4} \int_{\alpha_{i}}^{\alpha_{f}}d\alpha
\int_{\beta_{i}}^{1-\alpha} d\beta
\alpha\beta(1-\alpha-\beta)(10s^2-12s\widetilde{m}^2_c+3\widetilde{m}^4_c)
\nonumber \\
&&-\frac{m_s\langle\bar{s} g_s \sigma Gs\rangle}{128 \pi^4}
\int_{\alpha_{i}}^{\alpha_{f}}d\alpha \int_{\beta_{i}}^{1-\alpha}
d\beta \alpha\beta
\left[6(2s-\widetilde{m}^2_c)+s^2\delta(s-\widetilde{m}^2_c)\right]
\nonumber \\
&&+\frac{3m_sm_c^2\langle\bar{s}s\rangle}{32 \pi^4}
\int_{\alpha_{i}}^{\alpha_{f}}d\alpha \int_{\beta_{i}}^{1-\alpha}
d\beta(s-\widetilde{m}^2_c)\nonumber \\
&&+\frac{3m_s\langle\bar{s}g_s \sigma Gs\rangle}{128 \pi^4}
\int_{\alpha_{i}}^{\alpha_{f}}d\alpha\int_{\beta_{i}}^{1-\alpha}
d\beta\widetilde{m}^2_c-\frac{3m_sm_c^2\langle\bar{s}g_s \sigma
Gs\rangle}{128 \pi^4} \int_{\alpha_{i}}^{\alpha_{f}}d\alpha \, ,
\end{eqnarray}

\begin{eqnarray}
\rho_{\langle\bar{s}s\rangle^2}(s)&=&-\frac{m_c^2\langle\bar{s}s\rangle^2}{16
\pi^2} \int_{\alpha_{i}}^{\alpha_{f}}d\alpha
-\frac{\langle\bar{s}g_s \sigma Gs\rangle^2}{64
\pi^2M^2} \int_{\alpha_{i}}^{\alpha_{f}}d\alpha \left[ s-\frac{s^2}{2M^2}+\frac{\alpha(1-\alpha)s^3}{4M^4}\right] \nonumber \\
&&\delta(s-\widetilde{\widetilde{m}}_c^2)+\frac{m_c^2\langle\bar{s}s\rangle\langle\bar{s}g_s
\sigma Gs\rangle}{32 \pi^2} \int_{\alpha_{i}}^{\alpha_{f}}d\alpha
\left[1-\frac{1}{\alpha(1-\alpha)}+\frac{s}{M^2}
\right]\delta(s-\widetilde{\widetilde{m}}_c^2)
 \, , \nonumber\\
\end{eqnarray}

\begin{eqnarray}
\rho^A_{\langle GG\rangle}(s)&=&-\frac{m_c^2}{512 \pi^4}
\int_{\alpha_{i}}^{\alpha_{f}}d\alpha \int_{\beta_{i}}^{1-\alpha}
d\beta \left(\frac{\alpha}{\beta^2}+\frac{\beta}{\alpha^2}
\right)(1-\alpha-\beta)^3\left[2s-\widetilde{m}^2_c+\frac{s^2}{6}\delta(s-\widetilde{m}^2_c)\right]
\nonumber \\
 &&+\frac{3}{1024 \pi^4}
\int_{\alpha_{i}}^{\alpha_{f}}d\alpha \int_{\beta_{i}}^{1-\alpha}
d\beta(\alpha+\beta)(1-\alpha-\beta)^2(10s^2-12s\widetilde{m}^2_c+3\widetilde{m}^4_c)
\nonumber \\
&&-\frac{m_sm_c^2\langle\bar{s}s\rangle}{192 \pi^2}
\int_{\alpha_{i}}^{\alpha_{f}}d\alpha \int_{\beta_{i}}^{1-\alpha}
d\beta \left(\frac{\alpha}{\beta^2}+\frac{\beta}{\alpha^2}
\right)(1-\alpha-\beta)  \left[1+\frac{s}{M^2}+\frac{s^2}{2M^4}\right]\delta(s-\widetilde{m}^2_c)\nonumber \\
&&+\frac{m_sm_c^2\langle\bar{s}g_s \sigma Gs\rangle}{2304\pi^2M^6}
\int_{\alpha_{i}}^{\alpha_{f}}d\alpha \int_{\beta_{i}}^{1-\alpha}
d\beta \left(\frac{\alpha}{\beta^2}+\frac{\beta}{\alpha^2}
\right)s^2\delta(s-\widetilde{m}^2_c)\nonumber \\
&&-\frac{m_sm_c^4\langle\bar{s}s\rangle}{192\pi^2M^2}
\int_{\alpha_{i}}^{\alpha_{f}}d\alpha \int_{\beta_{i}}^{1-\alpha}
d\beta \left(\frac{1}{\alpha^3}+\frac{1}{\beta^3}
\right)\delta(s-\widetilde{m}^2_c)\nonumber \\
&&+\frac{m_sm_c^2\langle\bar{s}s\rangle}{64 \pi^2}
\int_{\alpha_{i}}^{\alpha_{f}}d\alpha \int_{\beta_{i}}^{1-\alpha}
d\beta \left(\frac{1}{\alpha^2}+\frac{1}{\beta^2}
\right)\delta(s-\widetilde{m}^2_c)\nonumber \\
&&+\frac{m_s\langle\bar{s}s\rangle}{128 \pi^2}
\int_{\alpha_{i}}^{\alpha_{f}}d\alpha \int_{\beta_{i}}^{1-\alpha}
d\beta(\alpha+\beta)\left[3+\left(2s+\frac{s^2}{2M^2}\right)\delta(s-\widetilde{m}^2_c)\right]
 \nonumber\\
 &&-\frac{m_sm_c^4\langle\bar{s}g_s \sigma Gs\rangle}{768\pi^2M^4}
\int_{\alpha_{i}}^{\alpha_{f}}d\alpha \int_{\beta_{i}}^{1-\alpha}
d\beta \left(\frac{1}{\alpha\beta^3}+\frac{1}{\alpha^3\beta}
\right)\delta(s-\widetilde{m}^2_c)\nonumber\\
 &&+\frac{m_sm_c^2\langle\bar{s}g_s \sigma Gs\rangle}{256\pi^2M^2}
\int_{\alpha_{i}}^{\alpha_{f}}d\alpha \int_{\beta_{i}}^{1-\alpha}
d\beta \left(\frac{1}{\alpha\beta^2}+\frac{1}{\alpha^2\beta}
\right)\delta(s-\widetilde{m}^2_c) \, ,
\end{eqnarray}

\begin{eqnarray}
\rho^B_{\langle
GG\rangle}(s)&=&\frac{m_c^4\langle\bar{s}s\rangle^2}{288M^4}
\int_{\alpha_{i}}^{\alpha_{f}}d\alpha
\left[\frac{1}{\alpha^3}+\frac{1}{(1-\alpha)^3}
\right]\delta(s-\widetilde{\widetilde{m}}_c^2)\nonumber \\
&&+\frac{m_sm_c^4\langle\bar{s}g_s \sigma Gs\rangle}{768\pi^2M^4}
\int_{\alpha_{i}}^{\alpha_{f}}d\alpha
\left[\frac{1}{\alpha^3}+\frac{1}{(1-\alpha)^3}
\right]\delta(s-\widetilde{\widetilde{m}}_c^2)\nonumber \\
 &&-\frac{m_c^2\langle\bar{s}s\rangle^2}{96M^2}
\int_{\alpha_{i}}^{\alpha_{f}}d\alpha
\left[\frac{1}{\alpha^2}+\frac{1}{(1-\alpha)^2}
\right]\delta(s-\widetilde{\widetilde{m}}_c^2)\nonumber \\
&&-\frac{m_sm_c^2\langle\bar{s}g_s \sigma Gs\rangle}{256\pi^2M^2}
\int_{\alpha_{i}}^{\alpha_{f}}d\alpha
\left[\frac{1}{\alpha^2}+\frac{1}{(1-\alpha)^2}
\right]\delta(s-\widetilde{\widetilde{m}}_c^2)\nonumber \\
&&-\frac{m_s\langle\bar{s}g_s \sigma Gs\rangle}{768\pi^2}
\int_{\alpha_{i}}^{\alpha_{f}}d\alpha
\left[1+\frac{s}{M^2}+\frac{s^2}{2M^4}\right]\delta(s-\widetilde{\widetilde{m}}_c^2)
\, ,
\end{eqnarray}

\begin{eqnarray}
\rho_{\langle
GGG\rangle}(s)&=&\frac{m_c^2}{4096\pi^6}\int_{\alpha_{i}}^{\alpha_{f}}d\alpha
\int_{\beta_{i}}^{1-\alpha} d\beta
\left(\frac{\alpha}{\beta^3}+\frac{\beta}{\alpha^3}
\right)(1-\alpha-\beta)^3\left[1+\left(\frac{2s}{3}+\frac{s^2}{6M^2}\right)\delta(s-\widetilde{m}^2_c)\right]
\nonumber \\
&&-\frac{1}{8192\pi^6}\int_{\alpha_{i}}^{\alpha_{f}}d\alpha
\int_{\beta_{i}}^{1-\alpha} d\beta
\left(\frac{\alpha}{\beta^2}+\frac{\beta}{\alpha^2}
\right)(1-\alpha-\beta)^3\left[2s-\widetilde{m}^2_c+\frac{s^2}{6}\delta(s-\widetilde{m}^2_c)\right]
\nonumber \\
&&+\frac{3}{8192\pi^6}\int_{\alpha_{i}}^{\alpha_{f}}d\alpha
\int_{\beta_{i}}^{1-\alpha} d\beta
\left(\frac{\alpha}{\beta}+\frac{\beta}{\alpha}
\right)(1-\alpha-\beta)^2\left[6s-3\widetilde{m}^2_c+\frac{s^2}{2}\delta(s-\widetilde{m}^2_c)\right]
\nonumber \\
&&-\frac{m_c^2}{4096\pi^6}\int_{\alpha_{i}}^{\alpha_{f}}d\alpha
\int_{\beta_{i}}^{1-\alpha} d\beta
\left(\frac{\alpha}{\beta^2}+\frac{\beta}{\alpha^2}
\right)(1-\alpha-\beta)^2\left[3+\left(2s+\frac{s^2}{2M^2}\right)\delta(s-\widetilde{m}^2_c)\right]
\, ,\nonumber \\
\end{eqnarray}
where $\alpha_{f}=\frac{1+\sqrt{1-\frac{4m_c^2}{s}}}{2}$,
$\alpha_{i}=\frac{1-\sqrt{1-\frac{4m_c^2}{s}}}{2}$,
$\beta_{i}=\frac{\alpha m_c^2}{\alpha s -m_c^2}$,
$\widetilde{m}_c^2=\frac{(\alpha+\beta)m_c^2}{\alpha\beta}$,
$\widetilde{\widetilde{m}}_c^2=\frac{m_c^2}{\alpha(1-\alpha)}$.


\begin{thebibliography}{99}

\bibitem{CDF0903}  T. Aaltonen et al,   Phys. Rev. Lett. {\bf 102} (2009) 242002.

\bibitem{Y4140-molecule} X. Liu and S. L. Zhu, Phys. Rev. {\bf D80} (2009) 017502;
 T. Branz, T. Gutsche and V. E. Lyubovitskij, Phys. Rev. {\bf D80} (2009) 054019;
  R. M. Albuquerque,  M. E. Bracco and M. Nielsen, Phys. Lett. {\bf B678} (2009) 186;
   G. J. Ding, Eur. Phys. J. {\bf C64} (2009) 297;
    J. R. Zhang and M. Q. Huang, J. Phys. {\bf G37} (2010) 025005;
     X. Liu and H. W. Ke,  Phys. Rev. {\bf D80} (2009) 034009;
      R. Molina and E. Oset, Phys. Rev. {\bf D80} (2009) 114013.


\bibitem{Y4140-tetraquark} Fl. Stancu, J. Phys. {\bf G37} (2010) 075017.
\bibitem{Drenska0902} N. V. Drenska, R. Faccini and A. D. Polosa,  Phys. Rev. {\bf D79} (2009) 077502.

\bibitem{Y4140-hybrid}  N. Mahajan, Phys. Lett. {\bf B679} (2009) 228.
\bibitem{Y4140-Wang} Z. G. Wang, Eur. Phys. J. {\bf C63} (2009) 115; Z. G. Wang, Z. C.
Liu and X. H. Zhang, Eur. Phys. J. {\bf C64} (2009) 373.


\bibitem{Y4140-cc} X. Liu, Phys. Lett. {\bf B680} (2009) 137; E. van Beveren and G. Rupp, arXiv:0906.2278.

\bibitem{Belle4350}  C. P. Shen et al, Phys. Rev. Lett. {\bf 104} (2010) 112004.

\bibitem{CDF1101} T. Aaltonen et al,  arXiv:1101.6058.

\bibitem{LiuLZ1011} X. Liu, Z. G. Luo and S. L. Zhu, arXiv:1011.1045.

\bibitem{HeLiu1102} J. He and X. Liu,  arXiv:1102.1127.

\bibitem{Finazzo1102} S. I. Finazzo, X. Liu and M. Nielsen, arXiv:1102.2347.

\bibitem{PDG} K. Nakamura et al, J. Phys. {\bf G37} (2010) 075021.

\bibitem{SVZ79}  M. A. Shifman, A. I. Vainshtein and V. I. Zakharov, Nucl. Phys. {\bf B147} (1979) 385.

\bibitem{Reinders85} L. J. Reinders, H. Rubinstein and S. Yazaki, Phys. Rept. {\bf 127} (1985) 1.

\bibitem{Mixing} T. Matsuki, T. Morii and K. Seo, Prog. Theor. Phys. {\bf 124} (2010) 285.


\bibitem{Lee2005} S. H. Lee, H. Kim and  Y. Kwon, Phys. Lett. {\bf B609} (2005)
252.

\bibitem{Nielsen0907} R. D. Matheus, F. S. Navarra, M. Nielsen and C. M.
Zanetti, Phys. Rev. {\bf D80} (2009) 056002.

\bibitem{Ioffe2005} B. L. Ioffe, Prog. Part. Nucl. Phys. {\bf 56} (2006) 232.

\bibitem{ColangeloReview} P. Colangelo and A. Khodjamirian, hep-ph/0010175.

\bibitem{Leinweber97} D. B. Leinweber, Annals Phys. {\bf 254} (1997) 328.


\bibitem{FluxTube} N. Isgur and J. E. Paton, Phys. Rev. {\bf D31} (1985) 2910;
 T. Barnes,  F. E. Close and E. S. Swanson, Phys. Rev. {\bf D52} (1995) 5242.

\bibitem{Latt} P. Lacock et al, Phys. Lett. {\bf B401} (1997) 308;
 C. W. Bernard et al, Phys. Rev. {\bf D56} (1997) 7039; P. Chen,   Phys. Rev. {\bf D64} (2001) 034509;
  X. Q. Luo and Y. Liu, Phys. Rev. {\bf D74} (2006) 034502.

\bibitem{QCD-string} Y. S. Kalashnikova and A. V. Nefediev, Phys. Rev. {\bf D77} (2008) 054025;
 V. Mathieu, Phys. Rev. {\bf D80} (2009) 014016.

 \bibitem{ccG} J. Govaerts,  L. J. Reinders, H. R. Rubinstein and J. Weyers,  Nucl. Phys. {\bf B258} (1985) 215.


\bibitem{Page1996} P. R. Page, Phys. Lett. {\bf B402} (1997) 183.

\bibitem{Page1997} P. R. Page,   Nucl. Phys. {\bf B495} (1997) 268.


\bibitem{X4350-Nielsen} R. M. Albuquerque, J. M. Dias and M. Nielsen, Phys. Lett. {\bf B690} (2010) 141.

\bibitem{X4350-Zhang} J. R. Zhang and M. Q. Huang, Commun. Theor. Phys. {\bf 54} (2010) 1075.

\bibitem{X4350-Ma} Y. L. Ma, Phys. Rev. {\bf D82} (2010) 015013.


\bibitem{X4350-Liu} X. Liu, Z. G. Luo and Z. F. Sun, Phys. Rev. Lett. {\bf 104} (2010) 122001.

\bibitem{X4350-Wang} Z. G. Wang, Phys. Lett. {\bf B690} (2010) 403.

\bibitem{X4350-Abud} M. Abud, F. Buccella and F. Tramontano, Phys. Rev. {\bf D81} (2010) 074018.


\bibitem{WangScalar} Z. G. Wang, Eur. Phys. J. {\bf C62} (2009) 375;
 Z. G. Wang, Phys. Rev. {\bf D79} (2009) 094027; Z. G. Wang, Eur. Phys. J. {\bf C67} (2010) 411.



\end{thebibliography}
\end{document}